# Electrochemical supercapacitor performance of SnO$_2$ quantum dots

Venkataramana Bonu,*[a] Bhavana Gupta,[a] Sharat Chandra,[b] Arindam Das,*[a] Sandip Dhara,[a] A. K. Tyagi[a]

[a]Surface and Nanoscience Division, [b]Materials Physics Division, Indira Gandhi Center for

Atomic Research, Kalpakkam-603102, India.

* Email: ramana9hcu@gmail.com, dasa@igcar.gov.in

**ABSTRACT**

Metal oxide nanostructures are widely used in energy applications like super capacitors and Li-ion battery. Smaller size nanocrystals show better stability, low ion diffusion time, higher-ion flux and low pulverization than bigger size nanocrystals during electrochemical operation. Studying the distinct properties of smaller size nanocrystals such as quantum dots (QDs) can improve the understanding on reasons behind the better performance and it will also help in using QDs or smaller size nanoparticles (NPs) more efficiently in different applications. Aqua stable pure SnO$_2$ QDs with compositional stability and high surface to volume ratio are studied as an electrochemical super capacitor material and compared with bigger size NPs of size 25 nm. Electron energy-loss spectroscopic study of the QDs revealed dominant role of surface over the bulk. Temperature dependent study of low frequency Raman mode and defect Raman mode of QDs indicated no apparent volume change in the SnO$_2$ QDs within the temperature range of 80-300 K. The specific capacitance of these high surface area and stable SnO$_2$ QDs has showed only 9% loss while increasing the scan rate from 20 mV/S to 500 mV/S. Capacitance loss for the QDs is less than 2% after 1000 cycles of charging discharging, whereas for the 25 nm SnO$_2$ NPs, the capacitance loss is 8% after 1000 cycles. Availability of excess open volume in QDs leading to no change in volume during the electro-chemical operation and good aqua stability is attributed to the better performance of QDs over bigger sized NPs.

**KEYWORDS**: Quantum dots, SnO$_2$, Super capacitor, Raman spectra of SnO$_2$.

1. Introduction

Semiconductor quantum dots (QDs) plays vital role in technological and biological applications. However for an effective use of QDs, it is very important to understand their distinct physical and chemical properties with respect to their bulk counterparts. Wide band gap $n$-type semiconducting $SnO_2$ nanostructures are well established for the use in technologically important applications like gas sensing, transparent electrodes, and as a catalyst [1-4]. It is also found that the $SnO_2$ is a potential material for much needed applications like solar cells [1,5,6], as anode materials in Li-ion battery [1,7-9], and electrochemical supercapacitors [11-13]. When compared with other metal oxides, $SnO_2$ holds high electric conductivity (21.1 $\Omega$cm), high theoretical capacity, low potential of Li ion intercalation, as well as superior electron mobility (100–200 $cm^2$/V.s). These properties made this material appealing for the energy storage applications [14,15]. However, $SnO_2$ or Sn-based anodes undergo a large volume changes of 200, 300% during the charging/discharging process for super capacitor and Li-ion battery applications, which leads to pulverization and loss of electrical contact between particles, subsequently, resulting in low capacity value and poor rate capability [16,17], To overcome this problem, structures are inter-connected to compensate the volume change [18]. Alternatively, nano-sized crystals with dense stacking is used to reduce the volume variation [19], along with mixture structures are formed with other inactive substances, e.g., $SnO_2$ with carbon [20,21], in which the inactive carbon behave as a confining buffer for the volume variations. However these techniques could not solve the problem completely. Because of the in-active carbon, the total capacitance value reduces. Smaller size nanocrystals can provide good stability, an increased interface for the interaction between $SnO_2$ and other species, leading to an enhanced performance in specific capacitance, power density [11,16]. Cui $et\ al.$ [11] reported pseudo capacitance of $SnO_2$ QDs by adding 20% acetylene black, which they attributed to Faradic reactions between $SnO_2$ QDs and the electrolyte. Shin *et al.*[22] reported maximum specific capacitance of 40.5 $\mu F/cm^2$ and 8.9% loss of specific capacitance after 1000 cycles for the pure hierarchical $SnO_2$ nanobranches without any carbon materials. $SnO_2$ QDs exhibit a typical behavior in comparison with their bulk counterparts due to high surface to volume ratio and excitonic confinement effects [23,24]. Thus, it is important to understand this distinct nature of QDs such that it can be used more effectively in applications. In the past, even though electrochemical studies were performed with the $SnO_2$ QDs, however, the reason behind the better performance of the smaller size NPs is not well understood.

In the present work we have studied electrochemical super capacitor performance of the pure 2.4 and 25 nm nanoparticles (NPs) without adding any additives like carbon materials to observe the pure material behavior. These nanostructures are studied for their stoichiometry by using electron energy-loss spectroscopy (EELS), specific surface area by using Brunauer–Emmett–Teller (BET) method and aqua stability by studying the time dependent absorption spectra and Zeta potential measurements. Possible change in the volume of the NPs is probed by utilizing the temperature dependent Raman spectroscopy. Density functional theory (DFT) calculations are also carried out to calculate the charge distribution for each atom and the relaxed surface of the atoms. The stoichiometry, aqua stability, volume changes of these 2.4 and 25 nm NPs are compared for their electrochemical super capacitor performance.

2. **Experimental section**

$SnO_2$ QDs were synthesized by soft chemical method. $NH_4OH$ (2M) drops were added to aqueous $SnCl_4$ (1M) solution under continuous magnetic stirring at 80 °C. Precipitated white gel was washed with mille-pore water (18.2 MΩcm) several times to remove the Cl ions. The washed gel was dried at 100 °C and then crushed with a rotor. The as-prepared material was further annealed in air atmosphere using horizontal quartz tube furnace for 1h at a temperature of 800 °C. Detailed synthesis process of NPs was reported in our earlier report [3].

Morphological and crystallographic information of the NPs were probed with the aid of field emission scanning electron microscopy (FESEM; Zeiss SUPRA 55) and transmission electron microscopy (TEM; Zeiss Libra 200), respectively. EELS studies were carried out using in-column second order corrected omega energy filter type spectrometer with an energy resolution of 0.7 eV in a high resolution transmission electron microscope (HRTEM, LIBRA 200FE Zeiss). Raman active vibrational modes were studied by using micro-Raman spectroscopy (InVia, Reinshaw) with 514.5 nm excitation of an $Ar^+$ laser with 1800 grooves/mm grating, and thermo electric cooled CCD detector in the back scattering configuration. The temperature dependent Raman spectra measurements were performed using an adiabatic stage (Linkam LNP95, UK). Optical absorption properties of the NPs were studied with the aid of UV-Vis spectroscopy (Avantes, Netherlands) in the range from 200 to 750 nm. For absorption study $SnO_2$ NPs of concentration 0.2 mg/ml were well dispersed in mille-pore water using ultrasonicator. Similar solution was used for the zeta potential (Malvern's Zetasizer-Nano) measurements. The specific surface area of the NPs powders were measured using the BET method (MONOSORB, MS16, Quantachrome, USA) employing nitrogen gas adsorption–desorption measurement at 77 K. In order to remove the moisture content on the surface, the NPs powders were degassed at 150 °C for 3 h in flowing argon prior to the measurement.

The charge distribution for each atom including surface atoms of the relaxed slab is calculated. The *ab-initio* DFT calculations are carried out using the VASP plane wave code [25] and PAW pseudo potentials with the PBE GGA flavor for the exchange and correlation interactions [26] on a $SnO_2$ (110) slab formed from the (110) planes in $SnO_2$ crystal comprising of two Sn-O tri-layers of a $2\sqrt{2}a_{SnO_2} \times 4c_{SnO_2}$ layer super cell. The surface of the slab is populated with H atoms attached to the out of the plane O atoms in order to simulate the gas molecules adsorbed on the surface due to the exposure to the atmosphere. The convergence of the total energy with respect to the energy cutoff and *k*-points is checked in both the crystal and the slab configurations and we have used energy cutoff of 400 eV and 10×10×16 k-points for the $SnO_2$ crystal and 5×5×1 k-points for the slab calculations. The final convergences for the self-consistent cycles are better than $10^{-6}$ eV for energy and $10^{-2}$ eV/Å for the individual forces in the ionic relaxation steps. The slab was formed from the $SnO_2$ crystal after obtaining the minimum energy configuration for the crystal. The extent of the vacuum region needed for the total energy convergence is also checked for the slab calculations. Final dimensions of the slab are 13.4871×12.9050×25.4352 Å. These dimensions are kept fixed for all the slab calculations and all the atom positions are relaxed further to minimize the inter-atomic forces. The 3D electron density distribution is obtained after the completion of the self consistent cycle. The ionic charges on the atoms have been analyzed using the Bader analysis [27] of the 3D electron density distribution.

For electrochemical characterization 2 mg of $SnO_2$ (both 2.4 and 25 nm) was dispersed in 100 μl of ethanol. After 15 min of sonication; 25 μl dispersion was taken and spread over FTO glass (covered with scotch tape) followed by dried under IR lamp. After drying, scotch tape was removed and selected area coated FTO glass was used for energy storage studies using three electrode system (Metrohm-Autolab model PGSTAT302N, containing Ag/AgCl as a reference electrode, Pt electrode as counter electrode and $SnO_2$ coated FTO glass as working electrode).

Electrolyte used for electrochemical studies was 0.5 M KOH. The FESEM images of the deposited thin films are given in the supplementary (Fig. S 1 and b). Unlike in the Fig. S1b, particles are not appearing in the Fig. S1a. This is due to the crystal size of the as-prepared sample is lesser than the resolution limit of the used FESEM.

3. **Results and discussion**

Low magnification TEM image (Fig. 1a) shows spherical as-prepared SnO$_2$ NPs. Average size of the as-prepared NPs (Fig. 1b) is around 2.4 nm, which is less than the Bohr radius of SnO$_2$ (2.7 nm). HRTEM image in Fig. 1c manifests crystalline and spherical SnO$_2$ QDs. Inset in Fig. 1c exhibits the zoomed image of single QD containing (110) crystalline plane of rutile tetragonal SnO$_2$ phase (JCPDS #41-1445) with a *d* spacing of ~3.45 nm. The selective area electron diffraction (SAED) pattern of the QDs shows continuous ring like pattern, which implies the availability of planes in all orientations (Fig. 1d). The blurring of the rings is due to the presence of nonuniform strain arising from the small sizes of the QDs in the samples.

Since the NPs radius is below the Bohr radius, the electron hole motion is not correlated. In that case, the effective band gap energy is given by eqn. 1 [28].

$$E_g^{\text{eff}} = E_g + \frac{\hbar^2 \pi^2}{2\mu r^2} - \frac{1.8e^2}{\varepsilon R} + \cdots \quad \ldots\ldots\ldots\ldots\ldots (1)$$

Here, $E_g$ is the bulk band gap energy (3.6 eV), *r* is the particle radius, *µ* is the effective reduced mass (0.27$m_e$ for SnO$_2$) [29], and the dielectric constant, ε, of SnO$_2$ is 14. Calculated band gap energy of SnO$_2$ QDs is ~4.4 eV from the eqn. 1. FESEM image of the SnO$_2$ NPs annealed at 800 °C is shown in Fig 2a. Average size of these NPs is measured to be ~ 25 nm (Fig. 2b). HRTEM image of the single 25 nm NP shows the crystalline (110) plane belong to rutile tetragonal SnO$_2$ with a *d* spacing value of 0.336 nm (Fig 2c). Spotty rings like SAED pattern (Fig. 2d) indicates better crystallinity than the as-prepared QDs.

Fig. 3 shows the EELS spectra of the 2.4 and 25 nm NPs. Peaks at 515 and 524 eV belong to the M$_5$ and M$_4$ delayed edges of Sn, respectively [30,31]. The O-K edge, which starts at 529 eV has two major peaks at 532.2 and 538.4 eV for the 25 nm NPs. The peaks are related to O 2*p* or O 2*p*–Sn 5*p* hybridized transitions and the seperation between them is ~6 eV (Fig. 3a). It indicates that the 25 nm NPs are in SnO$_2$ stoichimetric state [32]. Features above the 540 eV are ascribed to the multiple scattering resonances of SnO$_2$ [31]. Unlike in 25 nm NPs, the EELS spectrum of SnO$_2$ QDs exhibits single peak (Fig. 3). Qian *et al.*[33] studied the EELS spectrum (O-K edge) across the cross section of the transition metal oxide (Li[Li$_{1/6}$Ni$_{1/6}$Co$_{1/6}$Mn$_{1/2}$]O$_2$) layer from surface to bulk with a spatial resolution of 0.6 nm. In the bulk region they observed two distinct peaks for O-K edge similar to that observed for 25 nm NPs in our study. As the probe approached the surface of transition metal oxide layer, the first peak of the O-K edge at 532 eV gradually disappeared leaving single broad peak. The disappearance of the first peak is ascribed to the reduction of neighboring transition metal atoms and oxygen vacancy formation leading to non-stoichiometry. The two peaks in O-K edge of SnO$_2$ are related to the mixing of O 2*p* and Sn 5*p* orbitals [30-32]. In SnO$_2$, the O atom is surrounded by three Sn atoms in a triangular geometry. However, atoms on the surface lose this coordination. Additionally, the surface has a larger number of O vacancies and it is pre-dominant in QDs as surface to volume ratio value of QDs is ten times higher than that of 25 nm NPs. In our earlier report O vacancy related high intense photoluminescence of SnO$_2$ QDs was also reported [23]. The specific surface areas of the samples were calculated by using BET method. Specific surface areas of the 2.4 and 25 nm NPs are 158 and 11 m$^2$/g, respectively. The corresponding N$_2$ adsorption-desorption isotherms are given in the supplementary (Fig. S2 a and b). The high specific surface area of the 2.4 nm NPs is due to their 10 time high surface to volume ratio than the 25 nm NPs. These results are again supporting the EELS data.

Absorption spectra of the QDs (Fig. 4) and the 25 nm NPs (Fig. 5) were recorded with respect to time. Both the NPs were well dispersed on the first day of experiment by ultrasonication. By using Tauc's plot, the band gap of the QDs (inset Fig. 4) and the 25 nm NPs (supplementory Fig. S3) are measured to be 4.3 and 3.6 eV, respectively. The band gap value of QDs, measured using Tauc's plot for direct band gap materials [34], corroborates with the calculated band gap value of 4.4 eV using eqn. 1. The absorption spectrum of dispersed QDs and 25 nm NPs were collected in subsequent days by preserving them at 20 $^o$C. Except for the first day, the dispersed NPs were not ultrasonicated again, but shaken well before recording the absorption spectrum in later days. The absorption spectra of QDs did not alter in shape or in its intensity even after 30 days (Fig. 4). It indicates that the QDs are stable and do not agglomerate once they are dispersed. If the NPs agglomerates the absorption looks similar at all wavelengths [35]. The difference in absorption spectrum of agglomerated and dispersed NPs of size 4 nm were reported in our earlier report [35]. The absorption spectrum of the 25 nm NPs is seen to decrease in intensity within 12 hours (Fig. 5). This decrease in the absorption intensity reflects the precipitation of 25 nm NPs due to reagglomeration. Fu *et al.*[36] reported that the ZnO QDs remained stable in water due to surface hydroxyl (OH) groups. It is noteworthy that the presence of high density of OH¯ on the surface of the $SnO_2$ QDs were disseminated by using FTIR and Raman spectroscopic studies [23]. The 25 nm NPs obtained after annealing at high temperature did not show such widely covered presence of OH¯ groups on the surfaces [23]. The dominance of the surface in $SnO_2$ QDs is clear from the EELS spectrum. Bonded -OH groups allow help building van-der Walls bonding with the spread solvent. In addition, high surface to volume ratio in QDs provides buoyancy required for floating in the medium as oppose to 25 nm NPs which tend to settle due to gravitational force. Zeta potential measurements are carried out to find the stability of the NPs in aqua media (supplementary Fig. S4 a and b). Average zeta potential values are measured to be ~ +21 and +13 mV for the QDs and the 25 nm NPs, respectively. These values are further supporting the good stability of the QDs over the 25 nm NPs in aqua media.

We have estimated surface static charges induced by the surface -OH groups and by the in-plane O vacancy from DFT calculations to calculate the 3D electronic charge density of the $SnO_2$ slab. Then we have analyzed this charge density using the Bader analysis for understanding the stability of these nanostructures. The Sn and O atoms in the $SnO_2$ crystal show charges equal to the nominal charges of +4 and -2 for Sn and O respectively, signifying full charge transfer. While there is no change in the charges of the Sn and O atoms in the top most Sn-O layer just below the out-of-plane O atoms on the slab surface, the surface reconstruction due to the relaxation of the atom positions results in reduction of charges on the Sn atoms in the second Sn-O layer. Alternate rows of Sn atoms show charges of +3.22 and +3.87, while the in-plane O atoms show an excess charge of the order of -2.2. The bridging O atoms in between the two Sn-O layers do not develop any extra charge, but the top most out-of-plane O layer on the slab surface shows a reduction in the charge to -1.8 for the O atoms that are not bonded to H atoms. This shows that one of the possible mechanisms of how the static Coulomb charges can develop on the $SnO_2$ QD surface just because of the surface reconstruction and OH⁻ absorption, even in the absence of any surface defect and points to the polar nature of the surface. Defects like in-plane O vacancies cause a very large relaxation in the positions of all the Sn and O atoms in the slab. This rearrangement results in charge reduction on the in-plane Sn atoms near the O vacancy to +3.9, while other in-plane Sn atoms and those in the second Sn-O layer show much reduced charged states between +2.6 to +3.8. The O atoms in the top layer show a slight reduction in their charges to -1.9, while the other O atoms in the slab show changes in their charged state from -1.8 to -2.2. So the theoretical study demonstrates that the –OH groups helps in forming static charges which further cause for the stability of the NPs.

Temperature dependent Raman spectroscopic measurements were carried out on the QDs as well as 25 nm NPs (Figs. 6 and 7). Fig. 6 shows symmetry allowed Raman modes at 473, 632, 773 cm$^{-1}$ corresponding to $E_g$, $A_{1g}$, and $B_{2g}$ of rutile tetragonal $SnO_2$ [37]. Raman spectrum of 25 nm NPs at 80 K shows highest intensity for symmetry allowed Raman modes along with a blue shift. Whereas defect related forbidden modes ($A_{2u}$(TO), and $A_{2u}$(LO)) [37], which are prominent at high temperatures are not observed at 80 K (Fig. 6). Increase in the Raman intensity at 80 K may be a result of the decrease in bond length, where a reduction in the full width half maxima (FWHM) of the peak as well as increase in intensity can be observed [38,39]. The blue shift in the peak position with decreasing temperature is due to reduction of anharmonicity in the lattice [39]. Inset in Fig 6 shows the linearly fitted plot for $A_{1g}$ peak position vs. temperature. The slope of the linear fit that is temperature coefficient of the $A_{1g}$ peak found to be 0.0184 cm$^{-1}$/K.

Unlike 25 nm NPs the QDs exhibit a completely different Raman spectrum and an unusual behavior with respect to temperature. Fig. 7 shows the temperature dependent Raman spectra of $SnO_2$ QDs. There is hardly any signature from the Raman allowed modes of rutile tetragonal $SnO_2$. Other than the low frequency Raman mode, there is only one broad peak at 570 cm$^{-1}$ ($D$), which is characteristically observed for the $SnO_2$ QDs [23,37]. The $D$ peak shows complete temperature independence in the temperature range from 80 to 300 K. Unlike the defect modes in 25 nm NPs ($A_{2u}$(TO), and $A_{2u}$(LO)) the $D$ peak does not disappear or reduce in its intensity at low temperature of 80 K. Analogous to the EELS measurements, the Raman modes also depend on the coordination or local environment (leading to a specific local symmetry) of the atoms [39]. In our earlier report, we have showed that the *D* peak gradually disappears with increasing particle size [23]. The peak was assigned to the in-plane 'O' vacancies from the results of DFT calculations [40]. The reasons behind temperature independence of the *D* peak intensity may lie in the fact that the local co-ordination and hence the local symmetry of the atoms is not altered for the non-stoichiometric QDs (discussed in the EELS analysis Fig. 3) and also there is no significant anharmonicity arising in the temperature range of 80-300K. This is further supported by the size dependent low frequency Raman mode. The low frequency Raman scattering (LFRS) from the QDs exhibits an unusual behavior with respect to the change in temperature. Unlike symmetry allowed Raman modes in 25 nm NPs (Fig. 6), the LFRS mode in QDs increases in intensity with increasing temperature from 80 to 300 K. All the LFRS modes in Fig. 7 are fitted with Lorentzian function and the fitted parameters are depicted (supplementary information Fig. S5 and Table S1) to show that there are no appreciable change in the FWHM and peak positions. Wu *et al.*[41] observed a similar effect with respect to temperature for the $Ge_{0.54}Si_{0.46}$ nanocrystals embedded in $SiO_2$ matrix and have explained it with the help of the collective modes produced by the interference of Raman scattering from the individual QDs. They also found that the intensity of the LFRS mode increased with increasing temperature and the FWHM was independent of the temperature, similar to our observations. If the lattice parameter of QDs increases with increasing temperature then the total volume or size of the QDs also increases. Consequently the LFRS mode is supposed to show a red shift [23,42]. In contrast, the LFRS mode does not show any change in the position with increasing temperature indicating no change in volume for the QDs. It may be because of the fact that there is already a sufficient amount of open volume in the non-stoichiometric QDs to accommodate the temperature effect.

The EELS studies established the fact that the QDs are non-stoichiometric and, predominantly occupied with open volume. Time dependent UV-Vis studies and Zeta potential measurements revealed good aqua stability of the QDs. Temperature dependent Raman studies confirmed that there is no apparent volume change in the QDs with the increase in temperature indicating the presence of excess amount of open volume in the QDs. These properties are highly appealing for the super capacitor applications. The electrochemical super capacitive performances of the $SnO_2$ NPs of sizes 2.4 (QDs) and 25 nm were evaluated by a three-electrode testing configuration. The cyclic voltammetry (CV) curves of $SnO_2$ are casted over FTO as shown in Fig. 8 a and b. The shapes of the CV curves of the QDs at scan rates from 20 to 500 mV/s in the potential range of −0.2 to 0.5 V are closely similar to rectangular (Fig. 8b). There is no significant presence of both reduction and oxidation peaks. A pair of broad peak appeared at 0.12 (cathodic) and 0.135 (anodic). However, in case of the 25 nm NPs one pair of distinct redox couple appeared at 0.25 (cathodic) and 0.32 (anodic). Redox transformation of $SnO_2$ related to those peaks is shown in the following equation [43,44],

$$SnO_2 + K_a^+ + e_a^- \longleftrightarrow SnOOK_a$$

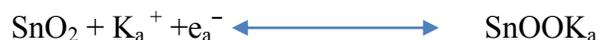

Shift in the peak potential towards higher potential side in case of 25 nm NPs of $SnO_2$ is due to the hindered redox/electronic transition in the $SnO_2$ film. However, in case of QDs there is an existence of a facile diffusion layer of ion due to its hydrous structure which retards the ionic flux towards the electrode. Opposite to that; in case of the 25 nm NPs large ionic flux from the bulk electrolyte is responsible for the formation of distinct redox active peak. Phenomenon of diffusion layer formation makes QDs superior in terms of columbic efficiency. Columbic efficiency was found to be 95 and 91 % for the $SnO_2$ NPs of size 2.4 and 25 nm respectively. Smaller particle size reduces the diffusion length and increases the reaction site for the ion due to its high surface area. This is certainly very important in regard to both capacitance and cyclic performance. Specific capacitance versus scan rate study has drawn immense attention in this direction (Fig. 8 c and d). Specific capacitance value for the QDs is 10 F/g at a scan rate of 20 mV/S and there is only 9% loss in the capacitance value after increasing the scan rate to 500 mV/S. Retention capacity of the 2.4 nm and the 25 nm particle is about 91 and 88% respectively (Fig. 8d). This result further concludes that the QDs are stable against the strain originated due to volume change at the time of electrochemical process. Similar effect was seen in the temperature dependent Raman spectroscopic studies. Such volume change creates a discontinuity between the particles subsequently responsible for reduced specific capacitance with scan rate as seen in the case of 25 nm NPs.

Galvanostatic charge discharge of the NPs was carried out to study the electrochemical behavior of materials with respect to time in a specific potential window in the same 0.5 M KOH solution. Potential window was kept of -0.2 to 0.5 V and -0.2 to 0.4 V for 2.4 and 25 nm NPs. Applied current density was 1 μA. Shape of the charge discharge curve in case of the 2.4 nm QDs was triangular symmetrical legs. However, symmetry got deviated in case of 25 nm NPs size (Fig. 8c). Ideal symmetric nature indicates the non-Faradaic process. Thus, Faradaic current component is enhanced in case of the 25 nm NPs in accordance to the cyclic voltammetric result. Potential window of the 25 nm NPs also reduced again due to enhanced Faradaic process above than 0.4 V potential. Cyclic stability of the QDs was tested in same electrolyte at $10^{-5}$ A applied current under the potential window of -0.2 to 0.5 V and -0.2 to 0.4 V for 2.4 and 25 nm $SnO_2$, respectively (Fig. 9). Thus, SnO$_2$ QDs posses extremely high electrochemical stability with retaining more than 98% of capacitance even after 1000 charge discharge cycles. Whereas, in case of the 25 nm NPs retainment of the capacitance is around 92% (Fig. 9). This further supports the superior stability of the SnO$_2$ QDs.

4. **Conclusion**

SnO$_2$ quantum dots (QDs) and bigger sized nanoparticles (NPs) of diameter 25 nm are studied as electrochemical super capacitor materials without adding any buffer materials. Dispersed QDs are found stable in water even after a month time. Electron energy-loss spectroscopic study of SnO$_2$ QDs showed significant absence of stoichiometry. There is no apparent volume change in the QDs within the temperature range of 80 to 300 K. Excess of open volume in the non-stoichiometric QDs is made responsible for it. There was only 9% loss in the capacitance value after increasing the scan rate from 20 to 500 mV/S. Capacitance loss for the QDs is less than 2% after 1000 cycles of charging discharging, whereas it is 8% for the 25 nm NPs. The new findings about the SnO$_2$ QDs can help in using it efficiently in different electronic and optoelectronic applications. The present studies can be extended to the other semiconductor QDs.

**Supporting Information**.

FESEM images of the deposited thin films of as-prepared and annealed NPs. Tauc's plot of the 25 nm NPs, Zeta potential measurements, Lorentzian fitting of the LFRS peaks in the Fig. 7 and Table containing Lorentzian fitted parameters of the LFRS mode in Fig. 7.

**Acknowledgements**

We acknowledge S. Amirthapandian, MPD, IGCAR for the EELS measurements and Dr. John Philip, SMARTS, IGCAR for the Zeta potential measurements.

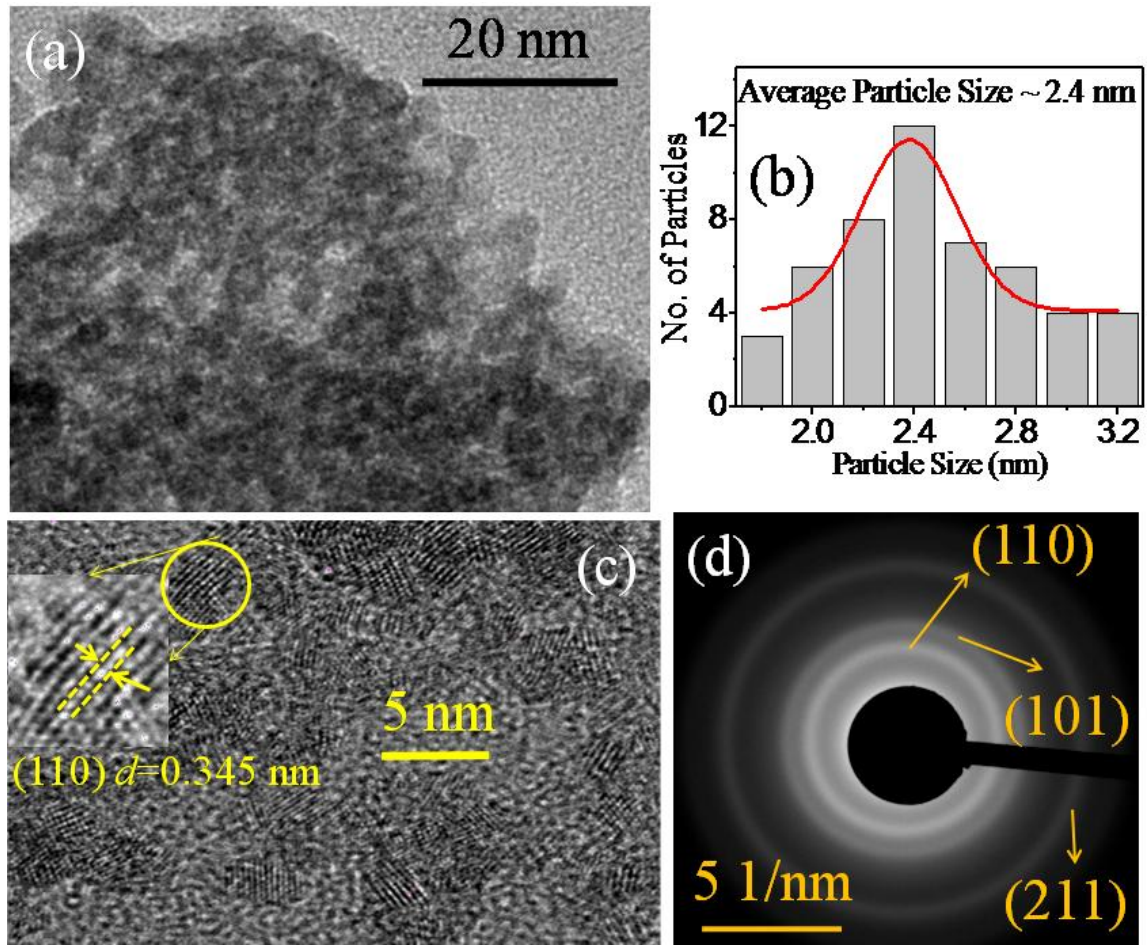

**Fig. 1.** TEM images of the as-prepared NPs (a) Low magnification TEM image, (b) Gaussian fitting of the size distribution, (C) HRTEM image, (d) SAED pattern.

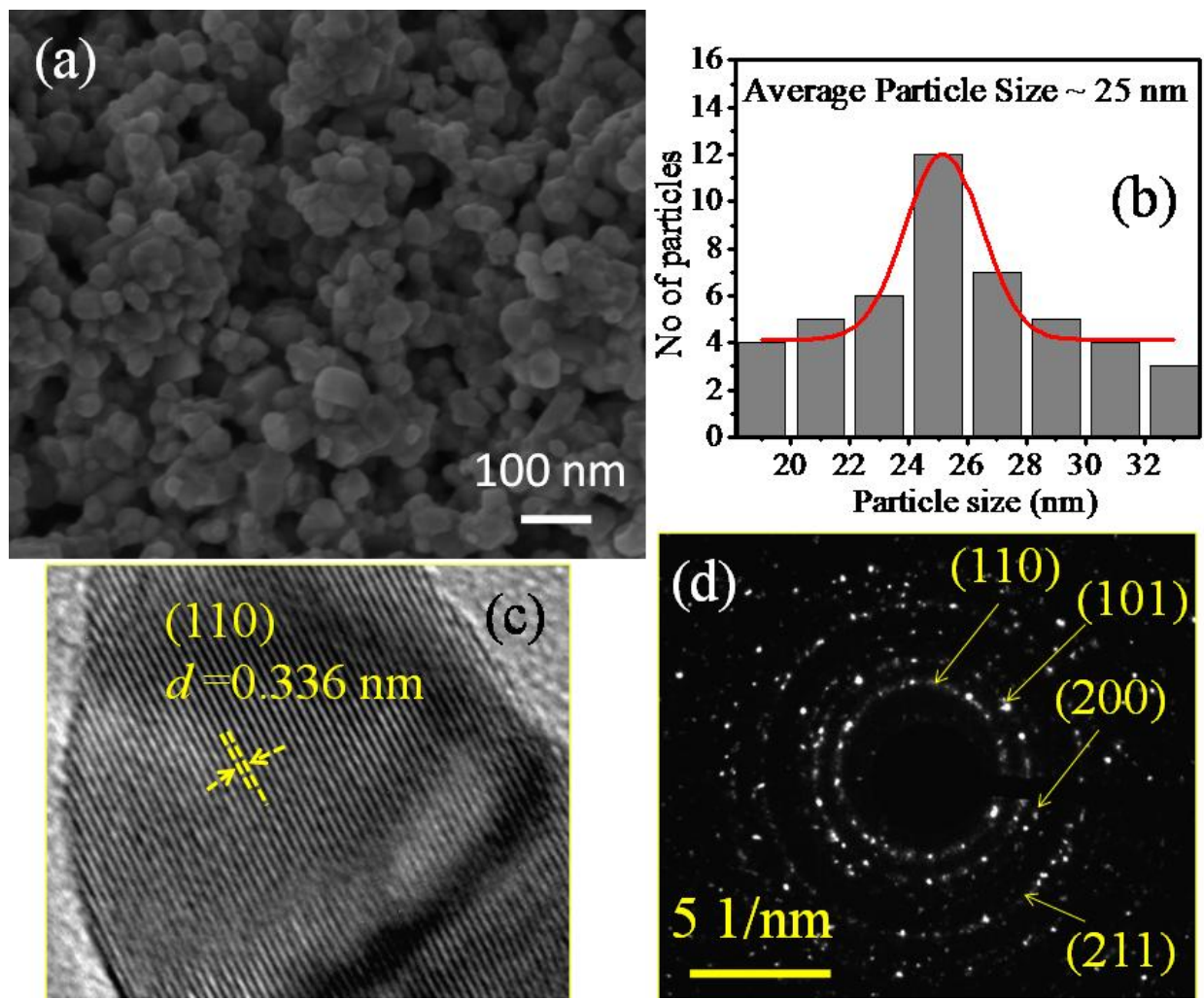

**Fig. 2** (A) FESEM images of the 800 °C annealed sample (b) Gaussian fitting of the size distribution, (C) HRTEM image, (d) SAED pattern the annealed NPs.

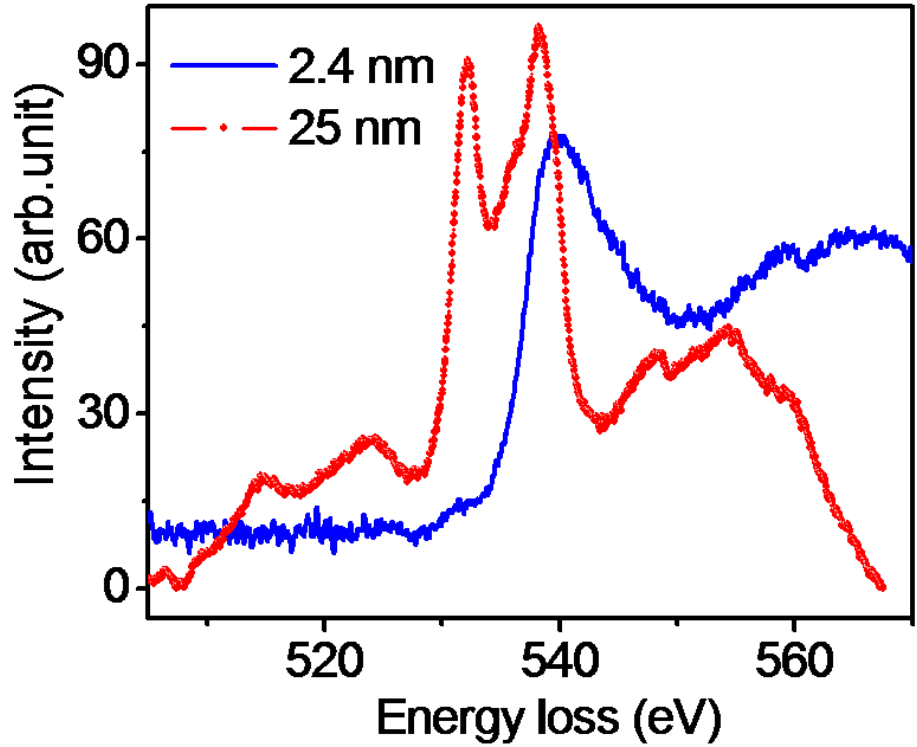

**Fig. 3.** EELS spectra of the SnO$_2$ QDs and the NPs of size 25 nm.

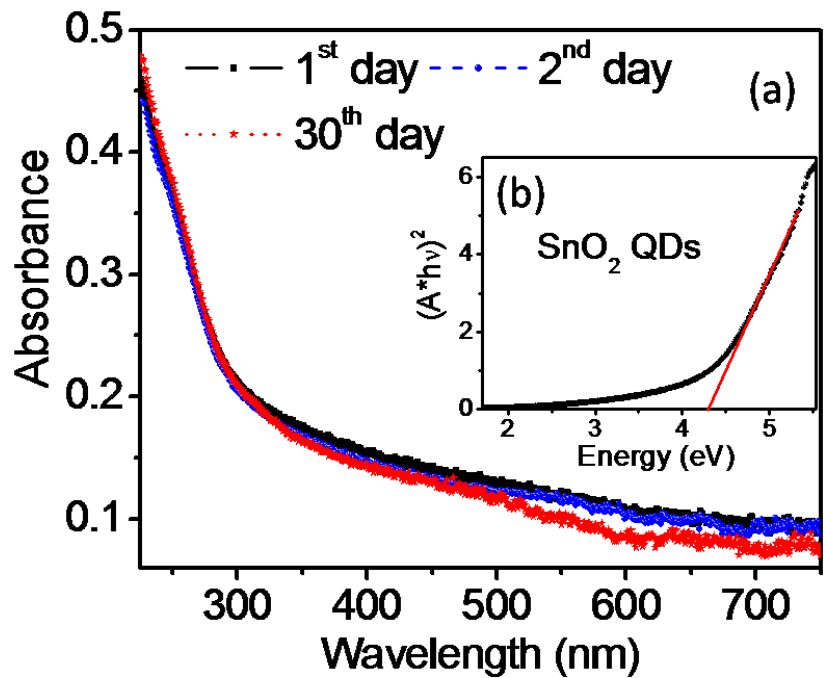

**Fig. 4.** (a) UV-Vis absorbance spectra of the QDs. (b) Tauc's plot of the QDs.

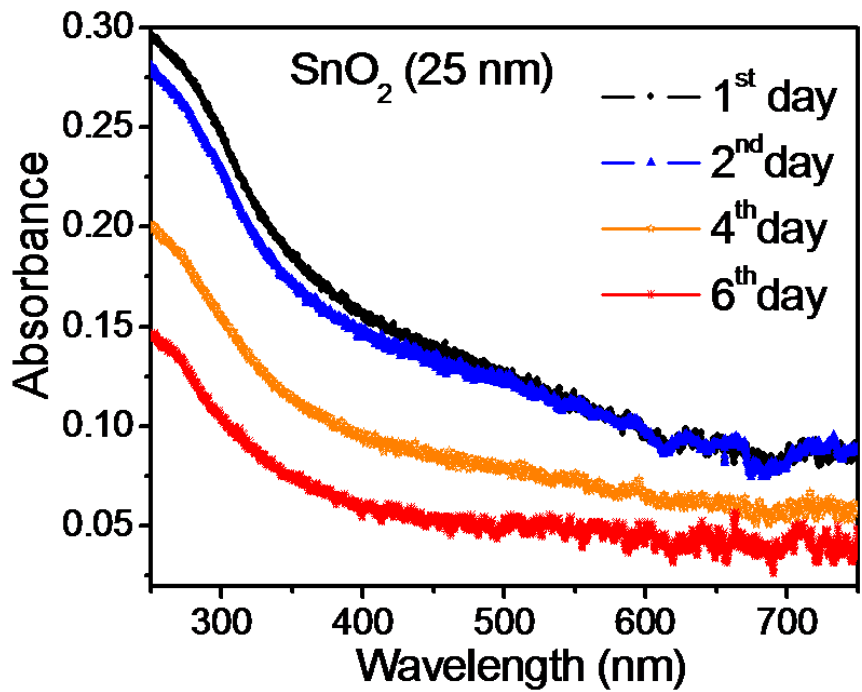

**Fig. 5.** UV-Vis absorbance spectra of the 25 nm NPs.

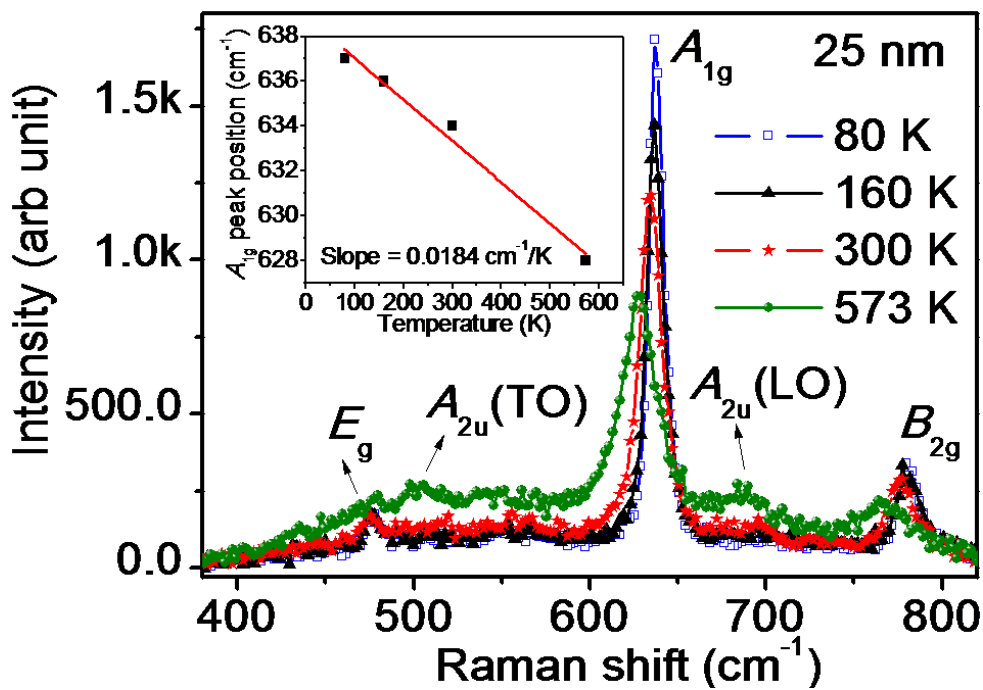

**Fig. 6.** Temperature dependent Raman spectra of the 25 nm NPs. Inset shows the plot between the $A_{1g}$ peak position and temperature.

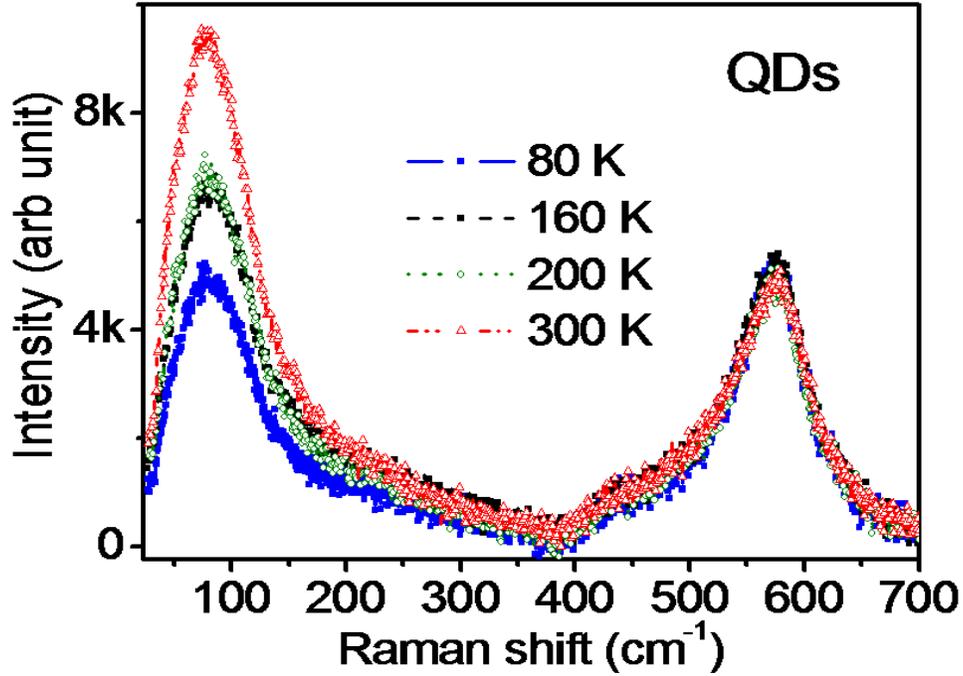

**Fig. 7.** Temperature dependent Raman spectra of the QDs.

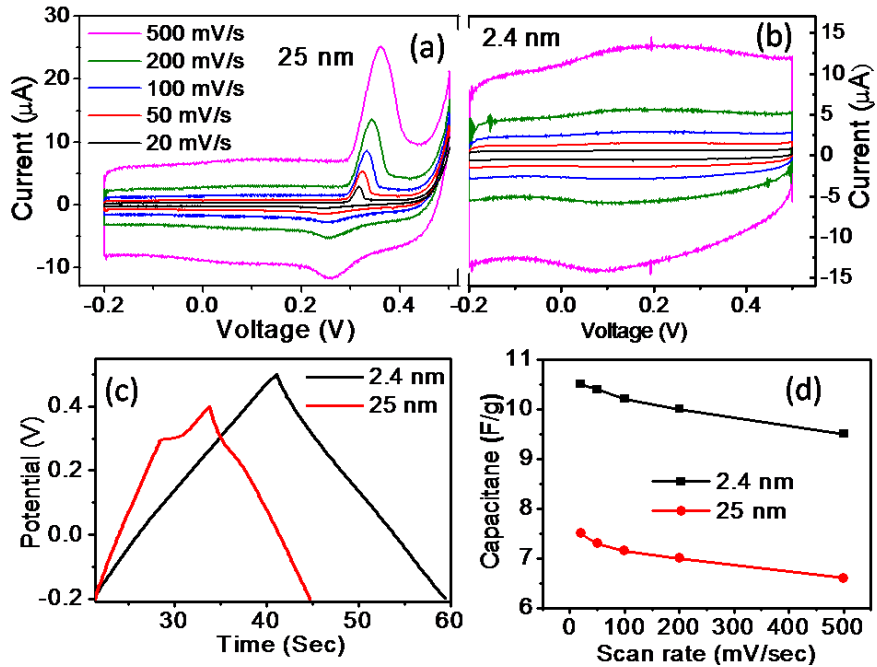

**Fig. 8.** Cyclic voltammetry of the (a) 25 nm NPs, and (b) 2.4 nm NPs. (c) Galvanostatic charge-discharge curves of the 2.4 and 25 nm NPs. (d) Capacitance of the 2.4 and 25 nm NPs with increasing scan rate.

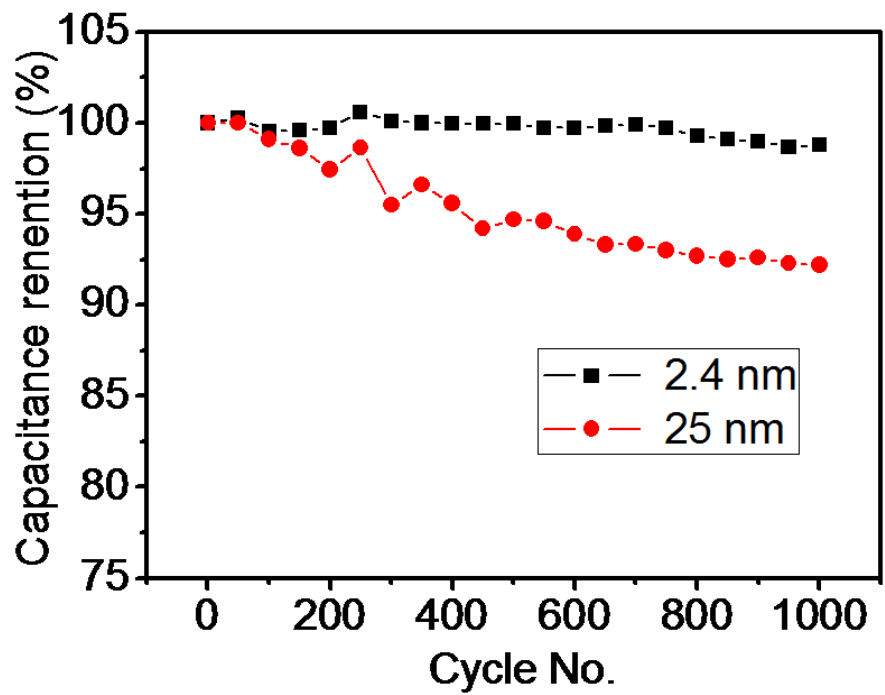

**Fig. 9.** Capacitance retention of the QDs and the 25 nm NPs with increasing cycle number at $10^{-5}$ A applied current.

# Supplementary

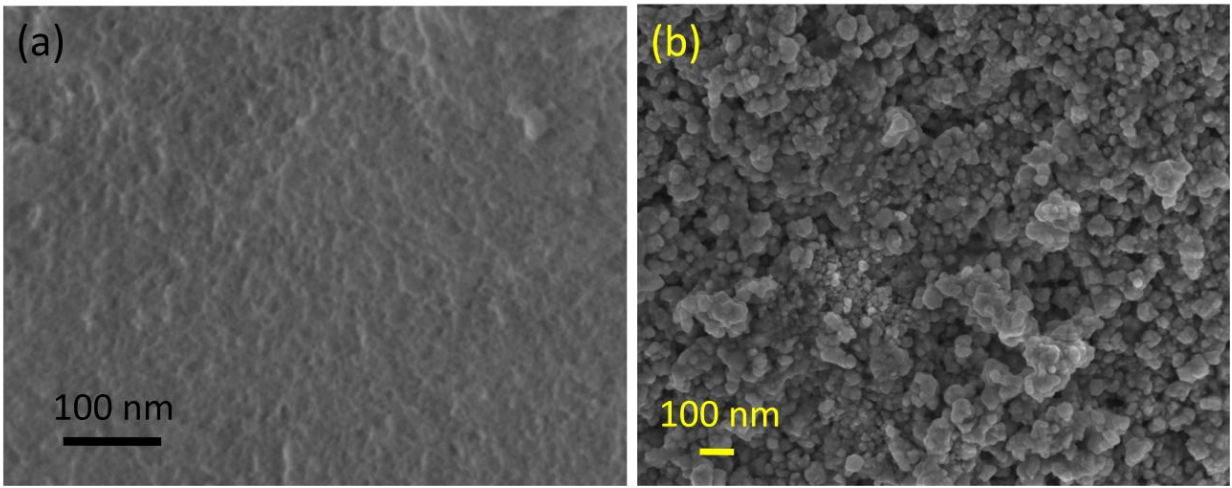

**Fig. S1** FESEM images of the deposited thin films. (a) As-prepared, (b) Annealed at 800 °C.

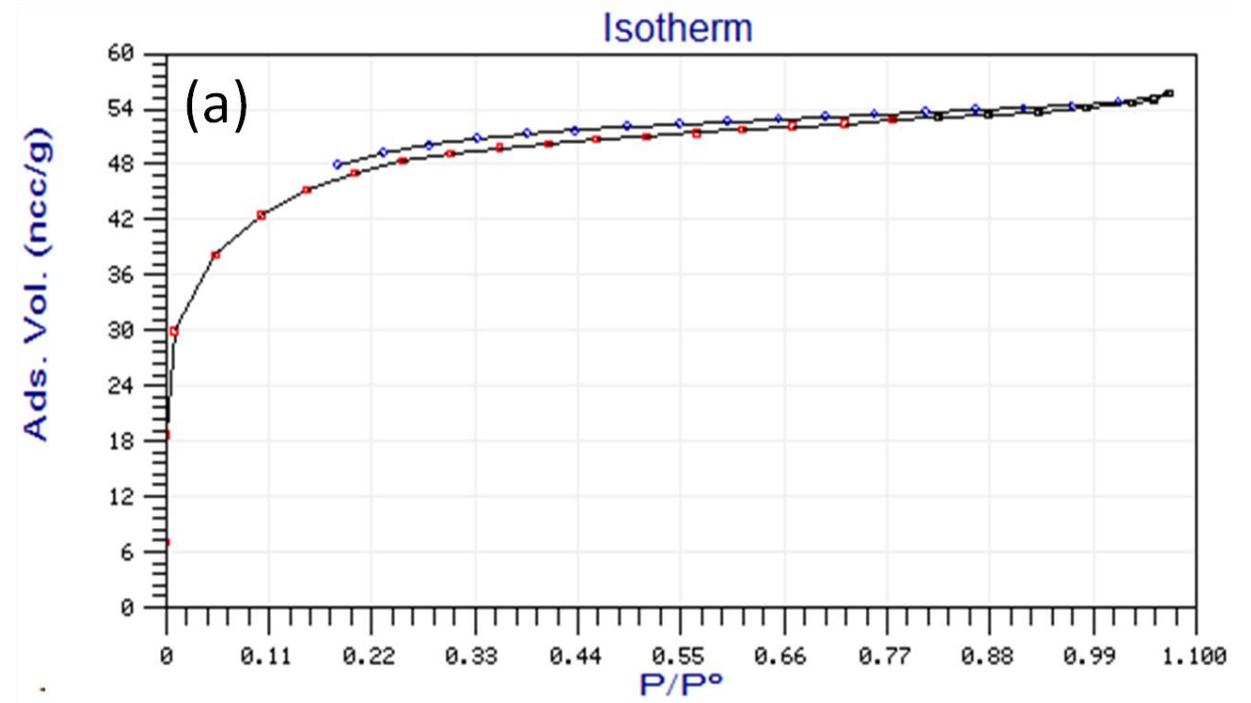
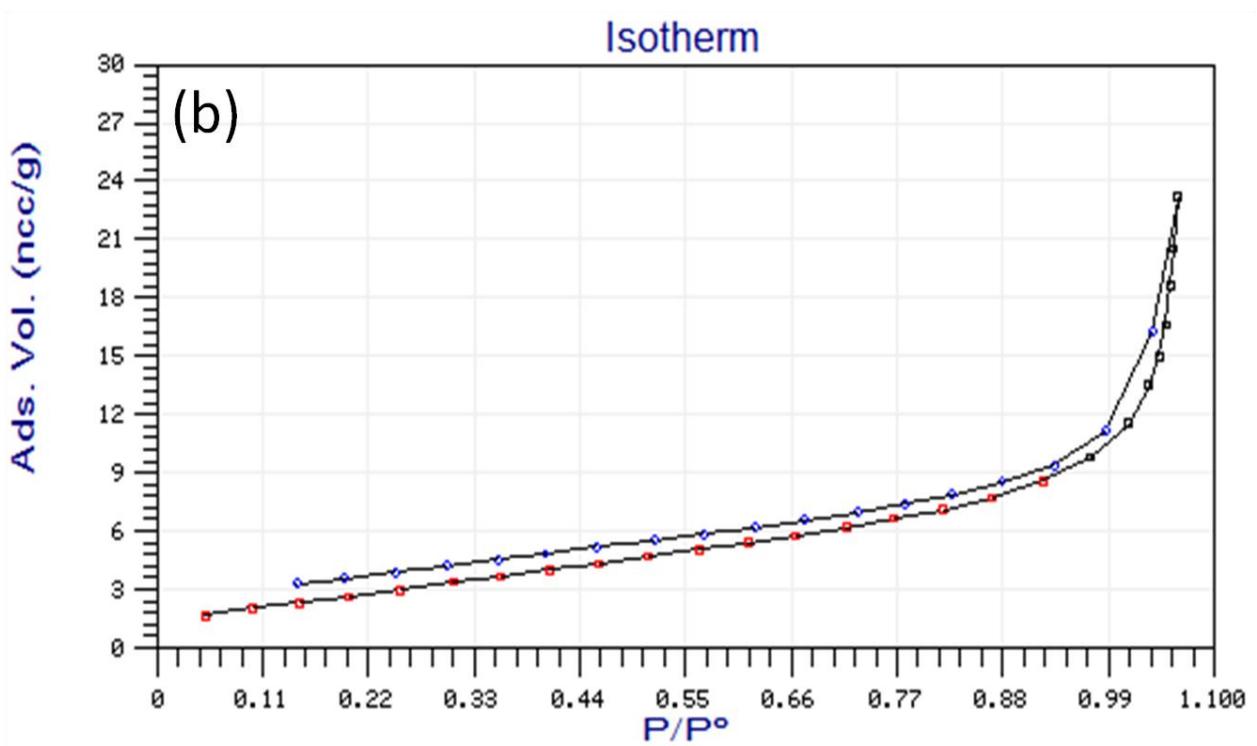

**Fig. S2** Nitrogen adsorption–desorption isotherms of (a) 2.4 nm NPs and (b) 25 nm NPs.

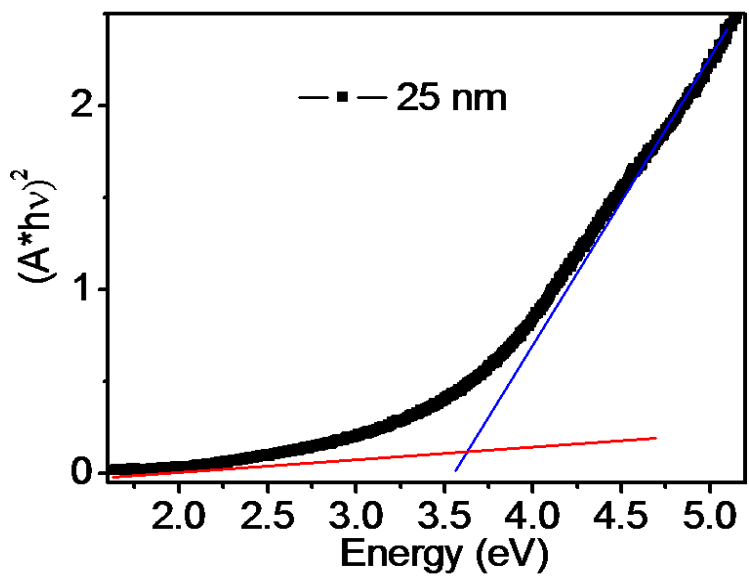

**Fig. S3** Tauc's plot of the 25 nm NPs. Measured band gap is 3.62 eV

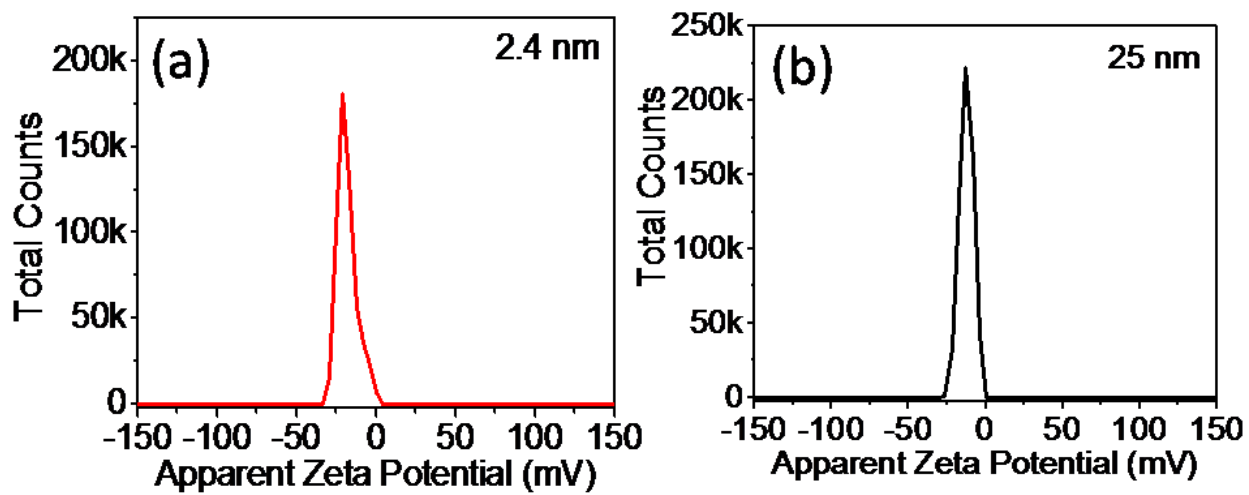

**Fig. S4** Zetapotential measurements of (a) 2.4 nm NPs and (b) 25 nm NPs.

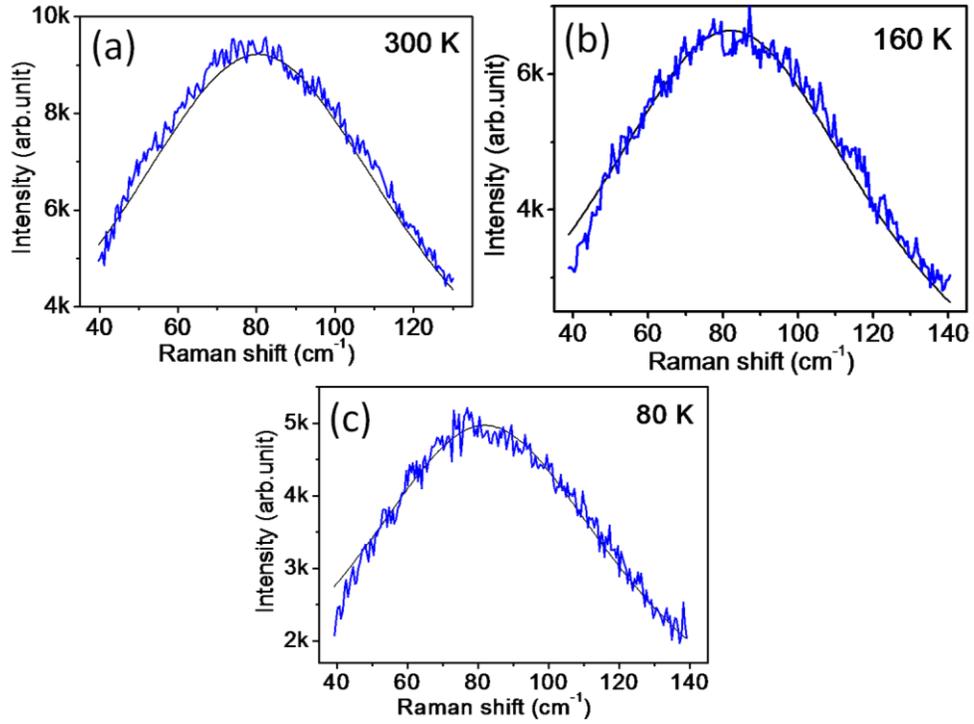

**Fig. S5** Lorentzian fitting of the LFRS peaks in the Fig. 6

Table S1. Lorentzian fitted parameters of the LFRS mode in Fig. S3.

| Temperature (°C) | Position(cm$^{-1}$) | FWHM (cm$^{-1}$) | Area(Square units |
|---|---|---|---|
| 80 K | 81 | 94 | 717243 |
| 160 K | 82 | 95 | 793833 |
| 300 K | 82 | 95 | 1223730 |